# Stimulated generation of deterministic platicon frequency microcombs

Hao Liu[1,+,*], Shu-Wei Huang[1,2,+], Wenting Wang[1], Jinghui Yang[1,3], Mingbin Yu[4], Dim-Lee Kwong[4], Pierre Colman[5], and Chee Wei Wong[1,*]

[1] Fang Lu Mesoscopic Optics and Quantum Electronics Laboratory, University of California Los Angeles, CA, USA

[2] Department of Electrical, Computer, and Energy Engineering, University of Colorado Boulder, CO, USA

[3] National Institute of Standards and Technology, Gaithersburg, MD, USA

[4] Institute of Microelectronics, Singapore, Singapore

[5] Université de Bourgogne Franche-Comté, ICB, UMR CNRS 6303, Dijon, France

[+] Equal contribution

[*] corresponding author: haoliu1991@ucla.edu; cheewei.wong@ucla.edu

**Dissipative Kerr soliton generation in chip-scale nonlinear resonators has recently observed remarkable advances, spanning from massively-parallel communications, self-referenced oscillators, to dual-comb spectroscopy. Often working in the anomalous dispersion regime, unique driving protocols and dispersion in these nonlinear resonators have been examined to achieve the soliton and soliton-like temporal pulse shapes and coherent frequency comb generation. The normal dispersion regime provides a complementary approach to bridge the nonlinear dynamical studies, including the possibility of square pulse formation with flat-top plateaus, or platicons. Here we report observations of square pulse formation in chip-scale frequency combs, through stimulated pumping at one free-spectral-range and in silicon nitride rings with +55 fs$^2$/mm normal group velocity dispersion. Tuning of the platicon frequency comb via a varied sideband modulation frequency is examined in both spectral and temporal measurements. Determined by second-harmonic auto-correlation and cross-correlation, we observe bright square platicon pulse of 17 ps pulsewidth on a 19 GHz flat frequency comb. With auxiliary-laser-assisted thermal stabilization, we surpass the thermal bistable dragging and extend the mode-locking access to narrower 2 ps platicon pulse states, supported by nonlinear dynamical modeling and boundary limit discussions.**

Over the past two decades, remarkable breakthroughs have been seen in the chip-scale frequency microcomb [1–5] studies with ultrahigh-Q microresonators [6–12], from table-top demonstrations [13,14] to small factor integrated platforms [15–17], thriving in both fundamental dynamics [14,18–26] and various applications, including spectroscopy [27], optical coherent tomography (OCT) [28,29], low noise radio frequency generation [30,31], frequency synthesis [15], distance ranging [32,33] and high-speed optical communication [34–36]. Most of these works are based on dissipative Kerr solitons (DKS) in anomalous group velocity dispersion (GVD) microresonators for self-referenced broadband optical frequency combs, which requires a delicate balance between loss and gain, as well as nonlinear phase and dispersion [37], and could limit the pulse energy if an ultrashort pulse and broadband spectrum are desired simultaneously. Since the material GVD of most platforms are normal in visible and near-infrared frequency range, the cavities are often engineered to achieve anomalous dispersion in pump resonances across a broad range [21,38,39]. Among these platforms, silicon nitride outstands due to its CMOS-compatible fabrication process, large Kerr nonlinearity, broad transparent window, low Raman nonlinearity and high-power handling capability [37]. The engineering of silicon nitride is commonly based on thick-nitride (>600 nm) waveguide that supports multiple transverse modes to achieve anomalous dispersion while maintaining high quality factors for low threshold frequency microcomb generation [17,40]. However, the inevitable coupling between different transverse mode families characteristically modulates the amplitude of the frequency microcomb spectrum and could detrimentally destabilize the dissipative Kerr soliton formation [41,42]. Tapered waveguide is then proposed to achieve anomalous GVD and single mode operation simultaneously [43,44], however, the operation for DKS remains non-trivial [17,45,46] with complex intracavity dynamics [22,47,48]. Furthermore, the ultralow loss thick-nitride waveguide fabrication needs special treatment [49], which is not offered in current commercial $Si_3N_4$ foundry process. Consequently, the adoption of DKS in standard PIC architectures through commercial foundries is still not possible [50]. Compared to the state-of-art ultralow-loss thick-nitride $Si_3N_4$ microresonator with 30-million $Q$ [51], recent demonstration of ultra-thin silicon nitride microresonators have achieved 260-million $Q$ with comparable free spectral range (FSR) [52], which significantly decreases the microcomb generation threshold. However, anomalous dispersion is forbidden in such scheme. Consequently, the study of novel methods to enhance normal dispersion frequency microcomb generation can simplify the microresonator frequency comb architecture and extend its operation into other frequency ranges.

Frequency microcomb generation in the normal GVD regime has recently been examined both theoretically and experimentally in a variety of platforms including crystalline resonators and integrated microresonators [21,53–56], as well as been demonstrated in many applications [35,57,58]. Its formation and nonlinear dynamics usually require shifted pump mode resonances [59], which can be achieved by avoided mode crossing caused by mode coupling between different mode families [60] or coupling between adjacent microresonators [61,62], or self-injection locking [50,58] to provide local anomalous dispersion [41,56]. Among these prior studies, solitonic bright pulse with unique flat-top square pulse shape – or the platicon – has increasingly drawn attention in numerical studies [59,63,64]. These theoretical modeling studies on the platicon have shown its pulsewidth can be continuously controlled in a broad range via the pump-resonance detuning (and hence the intracavity energy and nonlinear parametric gain). In addition, the conversion efficiency of pump power into comb power can be potentially higher in the platicon comb than in solitons, for the same GVD value [59]. Benefitting from its optical spectra's sharp edge and flat top features, the platicon can help increase the signal processing capabilities in optical domain for high-speed communication [65–67], with other applications in pulse shaping and amplification, nonlinear optical imaging, and production of high-brightness electron beams [68]. Although experimental demonstration of platicon frequency microcombs via self-injection locking has been recently demonstrated via self-injection locking [50] and pulse-pumping [69], the platicon generation via intensity-modulated pump [70], to the best of our knowledge, has not been demonstrated yet.

Here we experimentally demonstrate the generation of platicon frequency microcomb and its operating parameters in a chip-scale $Si_3N_4$ microresonator with normal dispersion, as a follow-up to our previous demonstration [71]. Via an intensity-modulated pump [70], a platicon generation approach is demonstrated such that sophisticated schemes for introducing pump mode shift or self-injection locking could be avoided. A platicon frequency comb with an 80-nm span (60 dB intensity roll-off) and clean comb spacing beat note is achieved. The platicon has a flattop pulse duration deterministically tunable from 2 to 17 ps, observed and confirmed through optical spectra, intensity autocorrelation and dual-comb cross-correlation measurements. Controls of the frequency comb optical spectrum through third-order dispersion (TOD), pump-resonance detuning, and modulation frequency are studied and characterized.

To understand and predict the nonlinear dynamics, we first start with a modified form of Lugiato-Lefever equation (LLE) which has an intensity-modulated external pump to numerically model the platicon generation:

$$T_R \frac{\partial}{\partial t}A(t,\tau) = \sqrt{\alpha_c}A_P - \left[\frac{\alpha_c+\alpha_p}{2} + j\delta - jL_{cav}\sum_{k\geq 2}\frac{\beta_k}{k!}\left(j\frac{\partial}{\partial \tau}\right)^k - j\gamma I(t,\tau)\right]A(t,\tau) \quad (1)$$

where $T_R$ is the round-trip time, $A(t,\tau)$ is the envelope function of the platicon, $t$ is the slow time corresponding to the evolution time over round trips, $\tau$ is the fast time describing the temporal structure of the wave, $A_P$ is the external pump, $\alpha_p$ is the propagation loss, $\alpha_c$ is the coupling loss, $\delta$ is the pump-resonance detuning, and $\omega_c$ and $\omega_p$ are the cavity resonance frequency and pump frequency, respectively. $\beta_k$ describes the dispersion coefficient ($\beta_2 > 0$ indicates normal GVD and $\beta_2 < 0$ indicates anomalous GVD), and we only consider second and third order dispersion in this case for simplicity. $\gamma = \frac{n_2\omega_0}{cA_{eff}}$ is the Kerr nonlinearity, in which $n_2$ is the nonlinear refractive index and $A_{eff}$ is the effective modal area of the pumping transverse mode. In our modified LLE, we assign $A_P = \sqrt{P\{1 + M\sin[\frac{2\pi t}{T_R}\left(\frac{\Delta}{\omega_{FSR}}+1\right)]\}}$ to describe the intensity-modulated external pump, where $P$ is the pump power without modulation, $M$ is the modulation depth ($0 \leq M \leq 1$), $\Delta = \omega_M - \omega_{FSR}$ is the deviation between the modulation frequency $\omega_M$ and the FSR $\omega_{FSR} = \frac{2\pi}{T_R}$ of the pumped cavity resonance. The simulation starts from vacuum noise and is run for $1.5 \times 10^5$ roundtrips until the solution reaches steady-state. The simulation is based on our FSR of 19 GHz with normal GVD of +55 fs$^2$/mm and negative TOD of -948 fs$^3$/mm.

The formation of platicon with intensity-modulated pump can be explained by wave-breaking theory [72–74] as detailed below. Figure 1a shows the 2D evolution map of platicon temporal profile against the non-dimensional cavity resonance detuning δ, with four snapshots of the 2D illustrated in Figure 1b. Figure 1b$_1$ (δ = -0.1) shows the sinusoidal envelope of the modulated pump input seeding the wave-breaking dynamics. Resulting from self-phase modulation, the pump ahead and behind the modulation minimum respectively experiences an instantaneous frequency up-shift and down-shift. In a normal GVD microresonator, such difference in the frequency shift leads to deceleration and acceleration of the pump ahead and behind the modulation minimum. Consequently, evolution of the initially sinusoidal envelope would be directed *outward* around modulation maxima and *inward* around modulation minima, leading to self-steepening around modulation minima [74], such as shown in Figure 1b$_2$ ($\delta$ = +0.0334). The wave is thus compressed and four-wave mixing (FWM) of the front of the pulse further broadens the comb spectrum, wherein the wave-breaking initialization occurs. This subsequently results in a bright and wide square pulse generation (pulse maxima spanning over a longer timescale), as shown in Figure 1b$_3$ ($\delta$ = +0.0598). With further red detuning ($\delta$ = +0.2531), the bright square pulse duration decreases down to 17 ps as shown in Figure 1b$_4$. Thus, after occurrence of the wave-breaking, the platicon pulsewidth shortens with increasing pump-cavity resonance red detuning while the peak intensity of the pulse increases as well.

The corresponding simulated frequency microcomb spectra are shown in the lower row of each intermediate states. We note that, in the platicon state, the comb lines have a somewhat flat plateau away from the pump wavelength with roughly uniform intensities, before gradually decreasing to the noise floor. In our numerical modeling plots, the *y*-axis range is chosen such that it is comparable to the experimental dynamic range in our optical spectrum analyzer (OSA). We put a 17-ps simulated platicon for better illustration of the square shape characteristics, however, the pulsewidth is theoretically controllable from tens of ps to sub-ps level, by simply varying the detuning. Since the 17 ps pulse has a square temporal structure, its computed autocorrelation is triangular as shown in Figure 1c, with the bottom width of the triangle twice that of the platicon square pulsewidth. The triangular structure in the autocorrelation helps distinguish if a platicon is generated in time-domain.

For our experiments, we utilize a single-mode $Si_3N_4$ microresonator nanofabricated in a CMOS-compatible foundry, with FSR of 19 GHz, *Q* factor of 1.2 million, and symmetrically top-bottom cladded with $SiO_2$. Through high-resolution coherent swept wavelength interferometry [53], the GVD is measured to be normal at 55 ± 2.5 fs$^2$/mm. Figure 1d subsequently shows the schematic setup for our measurements: the pump from a tunable external cavity diode laser (ECDL) is sent into an electro-optical modulator (EOM) with 20-GHz bandwidth. The modulation frequency, provided by a local oscillator (LO), is set at 19.548 GHz to closely match the single FSR of the microresonator at pump wavelength, and the modulation depth is set so that the sideband intensity is about 3 dB lower than the pump. After amplification by an erbium-doped fiber amplifier (EDFA), the polarization of the modulated is carefully controlled by a fiber polarization controller (PC) and a polarized beam splitter maximizes the intensity of the pump projected into the TM mode. Then the modulated pump with an on-chip power up to 1.4 W is launched into the microresonator. The output is collected by optical and electronic spectrum analyzers to measure the optical spectra, amplitude noises and radio frequency beat notes, respectively. Intensity autocorrelation and dual-comb cross-correlation measurements are conducted to analyze the time-domain performance of the platicon frequency comb.

We first experimentally study the evolution of the platicon frequency microcomb, with modulation frequency set at 19.548 GHz, close to 19-GHz single FSR. We red-tune the pump frequency in steps of 5 MHz to generate the platicon. The GVD and TOD measurement of the selected microresonator is shown in Figure 2a right inset. Figure 2a shows the triangular pump power transmission versus different pump-resonance detuning. States 1 (red) and 4 (black) are before and after the platicon generation, with their reference optical spectra (solely from the modulated pump) shown in the left inset of Figure 2a. The center pump is about 3 dB higher in power than the first sideband pair. As the pump laser is frequency tuned into the cavity resonance, the platicon frequency comb starts to evolve. At the beginning, cascaded FWM of the modulated pump leads to a weak growth of higher-order sidebands with characteristic rapid power decay for increasing mode numbers away from the pump. This is illustrated in the green curve of Figure 2b. With further tuning into the cavity resonance, the comb spectrum is dramatically broadened as shown in the blue curve of Figure 2b. The platicon spectrum shows the signature characteristic plateau and supported by our wave-breaking modeling shown earlier in Figure 1b. A typical single soliton spectrum envelope, clearly distinct from platicon is also included in Figure 2b as a light orange dashed line, which indicates that platicon could achieve higher comb line power at wavelengths away from the pump. The widest platicon comb spectrum generated by a single modulated pump is observed at the critical point of state 3, spanning over 80 nm (60-dB intensity roll-off) and matches our numerical modelled estimate of 82 nm (left inset of Figure 2b). Subsequently, with further increase of the pump wavelength, the platicon comb is lost and the spectrum drops back to the modulated pump line, as shown earlier in state 4.

To elucidate the underlying physics of the platicon formation, we next examined the impact of $\Delta$ on the platicon frequency comb spectrum. This is shown in Figures 3a to 3c (modeling) and Figures 3d to 3f (measurements). The platicon is robust under a range of $\Delta$ values, which effectively controls the spectral symmetry and intensity distribution of the comb lines. We consider the modelled platicon spectral characteristics prior to the measurements and for the three cases of zero third-order dispersion (TOD), negative TOD and positive TOD. Here, we use skewness from statistics for reference to describe the symmetry of platicon comb spectra (detailed in Supplementary Information Section II). In the first and conceptual case of zero TOD, our modified LLE modeling shows that $\Delta$ will affect the comb line distribution, hence the symmetry of the comb spectra. Since the comb spectrum is perfectly symmetric at $\Delta = 0$, the impacts of $\Delta$ on the comb line distribution is symmetrical around the zero point. (details in Supplementary Information Section II). We then examine the case of negative TOD at -1,000 $fs^3$/mm, matching our microresonator measurements (-948 $fs^3$/mm shown in top right inset of Figure 2a) and dispersion modelling (detailed in Supplementary Information Section I). When $\Delta$ is negative, the spanning of comb spectrum on the left-hand side is significantly extended, hence more comb lines are generated on the shorter wavelength side since the modulation frequency better matches the higher frequency FSR (shorter wavelength) due to positive GVD. This is shown in Figure 3a, for the case of $\Delta \approx$ -1,000 kHz. The comb span on the shorter wavelength side is almost 45-nm wider than the longer wavelength side, resulting in the platicon spectral asymmetry and more towards the blue-side.

Figure 3b shows that, when $\Delta \approx$ 0 kHz, the platicon microcomb is restored to a *symmetric* spectrum, spanning 80 nm. This arises because $\Delta$ balances out the effect of negative TOD. In Figure 3c, when $\Delta \approx$ 1,000 kHz, we see that the comb shape is asymmetric in an opposite manner (red-weighted) and has a distinctly narrower comb compared to the former two cases. This is due to the $\Delta$ and negative TOD both contributing to decreasing the phase matching bandwidth of the FWM. We note that, in each of the three $\Delta$ cases (Figure 3a to 3c), the general feature of the temporal square pulse is still captured in our simulations. Furthermore, we detailed the cases of zero and positive TOD, along with different $\Delta$, in Supplementary Information Section II, wherein the phenomenon is inverted and the platicon spectral asymmetry is at the opposite frequencies.

Figures 3d to 3f show our platicon measurements for three different modulation frequencies with $\Delta$ of -1,000 kHz, 0 kHz and 1,000 kHz. The general structure of the comb matches the numerical predictions remarkably. In Figure 3d ($\Delta$ = -1,000 kHz) we observe the blue-weighted platicon comb spectrum; in Figure 3e the symmetric comb spectra is observed. In Figure 3f ($\Delta$ = 1,000 kHz), the red-weighted platicon is achieved. We note that with a 2 MHz change in the modulation frequency, the platicon comb span is tuned by 20 nm and with a controllable blue-red weightage distribution of the platicon frequency comb. We also note that the modulation frequency should not deviate too much from the single FSR, otherwise, the broadband platicon comb is no longer observable (detailed in Supplementary Information Section II).

Next we examine the stability of the platicon comb and its time-domain characteristic of the platicon at stage 3. It is worth mentioning that since the comb spacing is intrinsically locked to the modulation frequency and there are no sub-comb families, the amplitude noise remains intrinsically low, and the electrical beat note remains clean with high signal-to-noise-ratio (SNR) throughout the whole evolution (detailed in Supplementary Information Section III). Figure 3h shows the measured power spectral density of the comb spacing beat note with a resolution bandwidth (RBW) of 1 kHz, centered at the $\approx$ 19.548 GHz modulation frequency. No other frequency components are observed. Differing from traditional dissipative Kerr solitons, the platicon is naturally mode-locked, avoiding modulation instability and high noise chaotic comb states. Figure 3i subsequently shows the platicon phase noise compared to the LO reference: at lower offset frequencies such as below 1 MHz, the platicon comb phase noise follows almost exactly the LO, with the platicon slightly worse than the LO in the flicker frequency ($1/f^3$) region below 180 Hz. However, starting from $\approx$ 1 MHz in our case, the platicon comb surpasses the phase noise character of the reference LO by up to 3 dB. Since this offset frequency is much lower than the cavity resonance linewidth ($\approx$ 157 MHz), this phase noise suppression should not be from the filtering effect of the resonator, but rather a phase noise low-pass filtering effect of the frequency comb, similar to [75].

The 60 dB SNR of the platicon beat note suggests that the frequency comb is phase-locked and potentially mode-locked. To verify the mode-locking, we next examine the intensity autocorrelation of the platicon. For a square pulse, the time-domain autocorrelation is triangular in structure. The measured autocorrelation of in Figure 4a shows a triangular shape with bottom width of 34 ps, indicating that a bright square pulse of 17 ps. Since this is a wide pulse, we implemented a dual-comb cross-correlation to depict the pulse shape. Sampled over 100 pulses, the dual-comb cross-correlated pulsewidth $\tau_a$ is gauged to be around 17 ps, and the repetition period ($\tau_a + \tau_b$) is observed to be 51 ps. And further detailed in Supplementary Information IV.

As previously mentioned, the pulsewidth of the platicon square pulse can be controlled by varying the pump-resonance detuning. Nonlinear thermal effect of the microresonator, however, introduces thermal bistable dragging when sweeping the pump across the resonance – this hinders us from accessing the effective red detuning side of the resonance, where narrower square pulses can exist. Hence we implemented an auxiliary-laser-assisted thermal stabilization method [46] to overcome the thermal bistable dragging (detailed in Supplementary Information Section V). Narrower deterministically tuned pulsewidths are successfully achieved. Figures 4c and 4d show the frequency comb spectra with pulsewidths of ≈ 4 ps and ≈ 2 ps respectively. The strong modulation of the comb spectrum arises from the Fourier nature of square pulse, which manifests itself as a sinc function in the frequency domain. The size of the main dome (first-minima spacing and bounded by the two vertical dashed blue lines as shown) uniquely corresponds to the pulsewidth $\tau_a$. Figure 4c shows a 0.54-THz spacing, which corresponds to a 4-ps square platicon. The frequency domain simulation (red envelope) matches the measurement almost exactly, which bridges the comb spectrum measurement to the corresponding time-domain simulation (inset figure). Figure 4d shows another achieved dual-pumped comb state with a 1.05-THz dome size, a Fourier shape that corresponds to the generation of a 2-ps square platicon. As previously discussed, the platicon pulse $\tau_a$ could be continuously changed by changing the detuning between the pump frequency and the resonance frequency. The inset of figure 4d shows the simulation of the pulse width vs. the detuning below 10 ps. The narrowest pulse width achievable with current simulation parameters is ≈ 616 fs. The inset shows that the pulse width becomes more sensitive to the detuning as the pulse width gets narrower.

In this work we demonstrate the platicon frequency comb generation in normal GVD microresonators, with a single-FSR intensity-modulated pumping scheme. Initiated from the wave-breaking dynamics, we analyze the influence of modulation frequency and pump-resonance detuning on the platicon frequency comb properties. A phase-locked and mode-locked frequency comb is observed and the bright square pulse with widths from 2 ps to 17 ps is depicted using dual-comb cross-correlation. We demonstrate the comb symmetry deterministic control with the sideband modulation frequency, together with the third-order dispersion, along with the beat note power spectral density and phase noise character. With auxiliary-laser-assisted thermal stabilization, we extend access of "the forbidden regions" of platicon generation, observing narrow square pulse generation. The stabilized platicon comb also simplifies estimation of the platicon pulse width via stable by bridging the platicon comb spectrum features to its pulsewidth. The nonlinear microresonator is a unique platform for the study of wave-breaking. This work has noteworthy influence on generating frequency microcombs in the normal dispersion regime and with applications such as an intensity-flattened spectral comb for high-rate optical communications, nonlinear optical imaging with fiber endoscopes, pulse shaping and amplification, dual-comb and Raman spectroscopy, and novel on-chip microwave synthesizers.

**Acknowledgements:** The authors thank the helpful discussions with Yongnan Li, James F. McMillan, Jinkang Lim, Abhinav Kumar Vinod and Qingsong Bai. We also thank Prof. Andrew Weiner for comments on initial discussions of this work.

**Author Contributions:** H.L. and S.W.H. designed the experiments and analyzed the data. H.L. and S.W.H. performed the experiments. S.W.H. and J.Y. designed the resonator, and M.Y. and D.L.K. fabricated the microresonator. W.W. helped on the phase noise measurement. H.L. and C.W.W. prepared the manuscript. All authors contributed to discussion and revision of the manuscript.

**Funding:** This material is based upon work supported by the Office of Naval Research (N00014-16-1-2094), the National Science Foundation (1741707, 1810506, and 1824568), and the Air Force Office of Scientific Research (FA9550-15-1-0081; Huang).

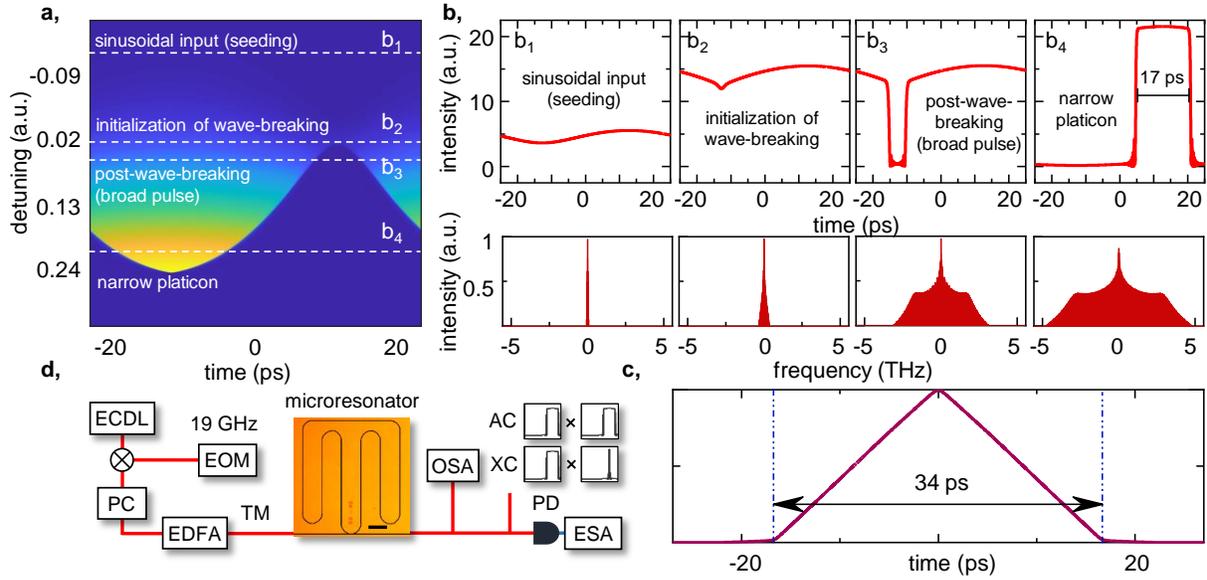

**Figure 1 | Platicon pulse generation in normal-dispersion frequency combs: operating regimes, numerical modeling and modulated pump experimental setup. (a)** 2D evolution map of platicon temporal profile as a function of intracavity fast time and detuning. Four characteristic stages are selected to show the details of the evolution in panel **b**. **(b)** Upper panels: temporal profiles of the platicon at different evolution stages, lower panels: frequency spectra corresponding each state. **b$_1$:** sinusoidal input (seeding); **b$_2$:** initialization of wave-breaking. Self-steepening shows that wave breaking is about to happen; **b$_3$:** post-wave-breaking. Wave breaking happens, and a broad square pulse is generated; **b$_4$:** a shorter narrow platicon is generated for increasing red detuning such as at δ = +0.2531. **(c)** The simulated autocorrelation of the square pulse shown in Figure 1b$_4$. The width of the triangle is twice of the width of the square pulse. **(d)** Schematic setup of platicon generation. The cw laser from the ECDL is first modulated by an EOM, and the modulation frequency is chosen to match the single FSR (≈ 19.547 GHz). After amplification by an EDFA, the modulated pump is launched into the single mode microresonator, and the pump frequency is slowly red-tuned to generate the platicon. A polarization controller (PC) and a polarized beam splitter (not shown in the diagram) are utilized to optimize the intensity of TM polarization for TM mode operation. A 20-GHz high-speed photodetector (PD) is used to measure the amplitude and phase noises. An OSA is used to map the platicon spectrum. Then the time domain dynamics is examined by both auto-correlation (AC) and dual-comb cross-correlation (XC). The actual single mode microresonator is shown in the middle micrograph, whose straight waveguide is tapered from 2.5 µm to 1 µm to maintain high quality factor. The curved regions are 1 µm in width to maintain our single-mode frequency comb operation. Scale bar: 200 µm.

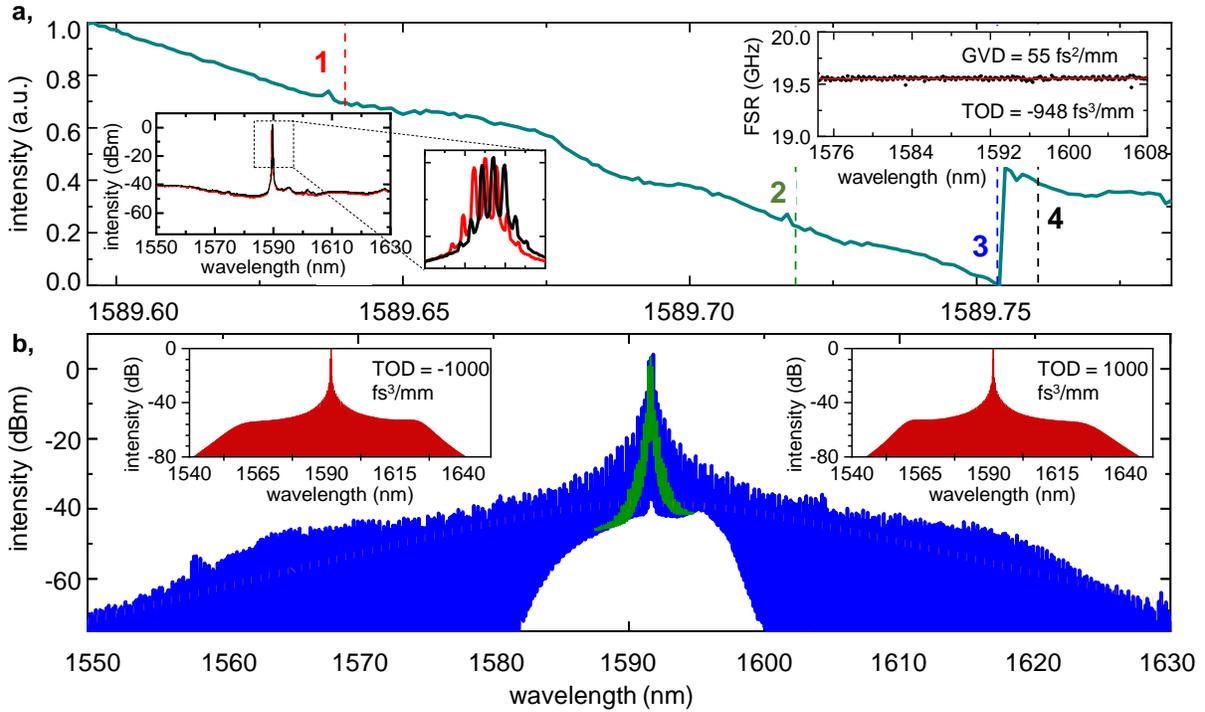

**Figure 2 | Dynamic detuning evolution and spectra of a modulated-pump platicon frequency comb.** **(a)** Pump transmission vs. detuning. The comb evolves to its critical point at state **3** (blue). Left inset: optical spectra of states **1** (red) & **4** (black), in which there are only modulated pump spectra. The first pair of sidebands are about 3 dB lower than main pump. Right inset: FSR measurement of the microresonator. The fitted GVD is positive at 55 $fs^2$/mm, and the fitted TOD is negative at -948 $fs^3$/mm. **(b)** Comb spectra at state **2** & **3**. The comb just starts to evolve due to spontaneous FWM at state **2** (green curve) and the spectrum shape is more like a triangle other than the frequency spectrum of a platicon. At state **3**, the frequency comb spectrum (blue curve) is at the critical point, spanning 80 nm. The shape coincides with the simulation, shown in the upper left inset. The upper right inset is comb spectrum simulation with positive TOD, as comparison. A typical single soliton comb spectrum envelope is included in dash pink line, indicating that platicon has higher comb line power at wavelength away from the pump.

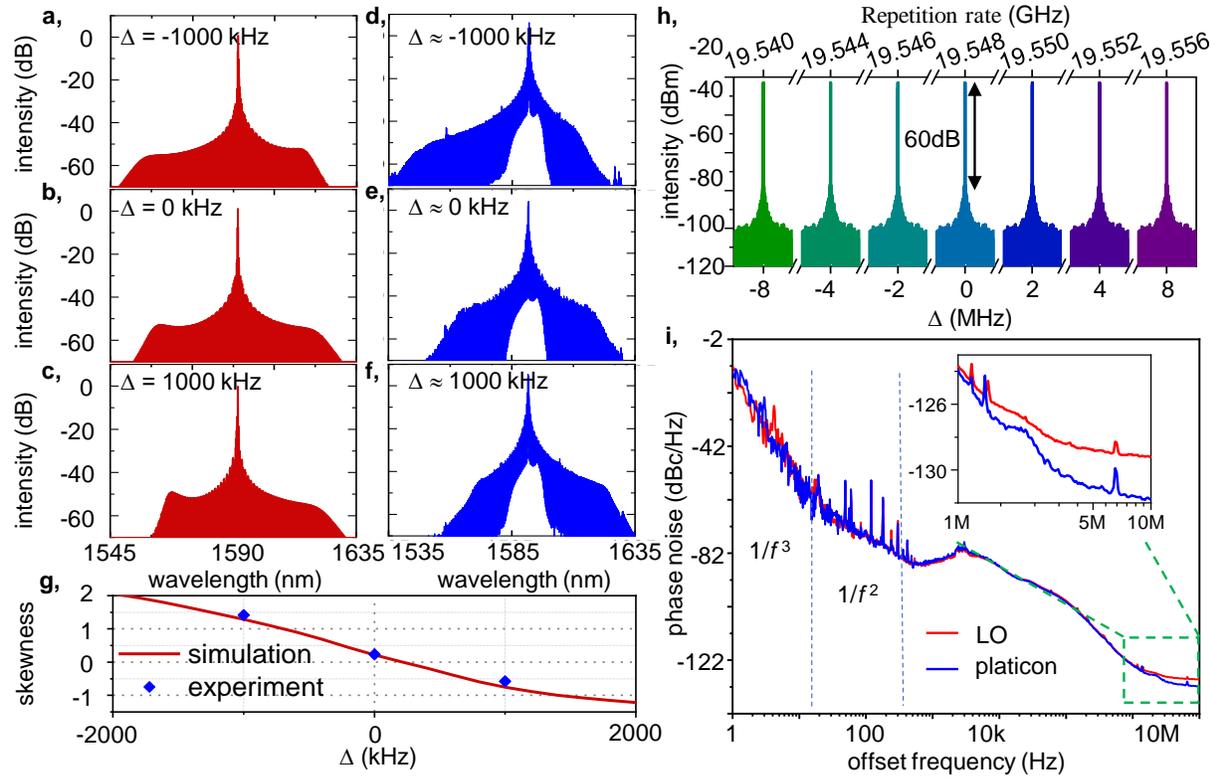

**Figure 3 | Deterministic platicon comb formation controlled by the sideband modulation frequency. (a)-(c)** Simulated frequency comb spectra with different modulation frequencies under a TOD of -1,000 fs$^3$/mm. **a:** $\Delta$ = -1000 kHz, **b:** $\Delta$ = 0 kHz, **c:** $\Delta$ = 1000 kHz. **(d)-(f)** Experimental frequency comb spectra with different modulation frequencies. **d:** 19.547 GHz ($\Delta \approx$ -1000 kHz), **e:** 19.548 GHz ($\Delta \approx$ 0 kHz), **f:** 19.549 GHz ($\Delta \approx$ 1000 kHz). All comb spectra span around 80 nm. **(g)** Comb skewness versus sideband modulation frequency $\Delta$. **(h)** heterodyne beat note of the platicon frequency microcomb at different modulation frequency. The *x*-axis is the offset frequency with respect to the FSR of pump mode. A 60-dB SNR is observed for each case. The RBW is 1 kHz. **(i)** phase noise measurements of the LO and platicon comb. The phase noise of platicon comb almost exactly follows the phase noise of the LO in the low frequency range, except in the 1/$f^3$ range, the platicon is slightly better. But starting from 1 MHz, the platicon comb outperforms the LO, and suppresses the phase noise by up to 3 dB.

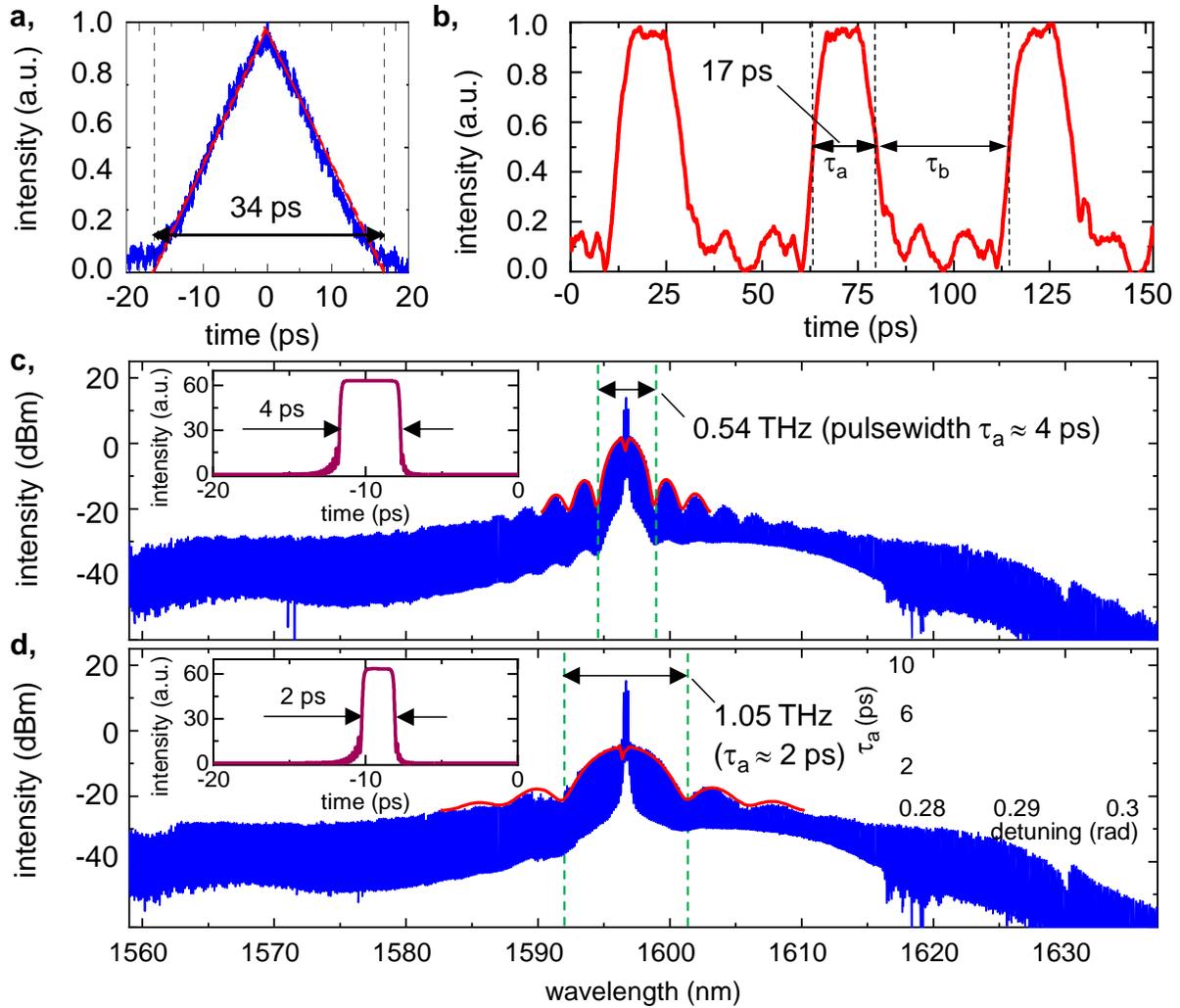

**Figure 4 | Pulsewidth characterization of platicon square pulses. (a)** Time-domain autocorrelation of the platicon comb. The bottom width of 34 ps of the triangle shape indicates a square bright pulse of 17 ps. **(b)** Cross-correlation of the platicon pulse frequency comb. A 17 ps square bright pulse is directly observed. **(c)** Comb spectrum of a 4 ps platicon square pulse. The spacing between the two first-minima is about 0.54 THz. The experimental measurement matches well with simulation platicon comb spectrum (red curve). The inset is the time-domain square pulse simulation corresponding to the red curve, which proves a 4-ps square pulse generation. **(d)** Comb spectrum of a 2 ps platicon square pulse. The simulation of both frequency- and time-domain shows a 2-ps square pulse generation. Left inset is the time-domain profile. Right inset is a summary of the platicon pulsewidths versus δ, which indicates that narrow pulsewidth down to hundreds of fs could be achieved.

# Supplementary Information

This Supplementary Information consists of the below sections:

**I. Dispersion design and characterization of the $Si_3N_4$ microresonator**

**II. Impact of Δ on the platicon generation with zero and positive TOD**

**III. Noise performance characterization**

**IV. Dual-comb cross-correlation**

**V. Auxiliary-laser-assisted thermal stabilization**

## I. Dispersion design and characterization of the $Si_3N_4$ microresonator

Figure S1 shows the calculated GVD and TOD of fundamental TM mode in different waveguide width and tapered 19 GHz $Si_3N_4$ microresonator based on effective index simulation via COMSOL. The 19 GHz microresonator has 74 % tapered straight waveguide and 26% bending waveguide. The straight tapered waveguide starts from 1 μm width to 2.5 μm width and then back to 1 μm width to seamlessly connect the bending waveguide. And the overall GVD for the tapered microresonator is around 50 $fs^2$/mm, which is close to our measurement, while the overall TOD is about -1,000 $fs^3$/mm at pumping wavelength.

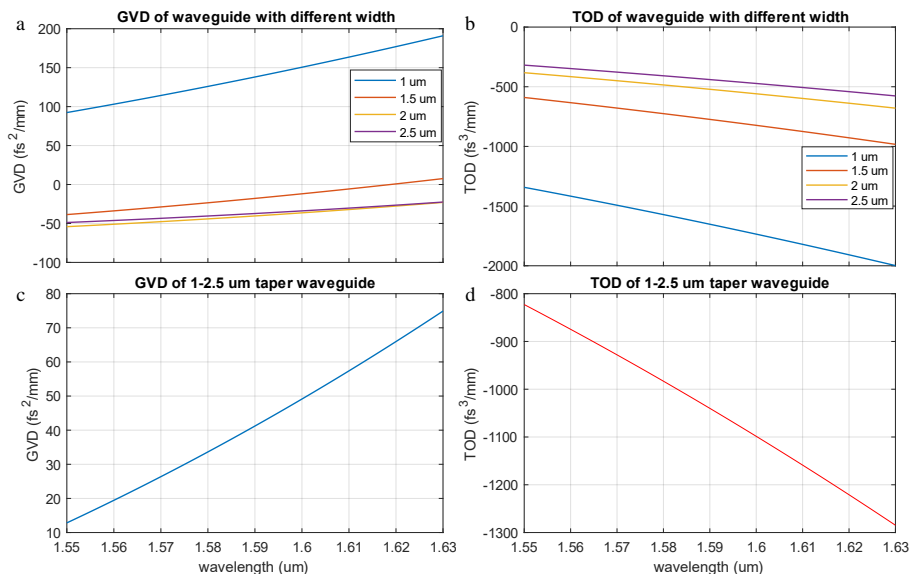

**Figure S1 | GVD and TOD simulation of the tapered $Si_3N_4$ microresonator.** (a) GVD modelling for different waveguide width, at fixed waveguide height of 800 nm. (b) TOD modelling for different waveguide width, at fixed

waveguide height of 800 nm. **(c)** Net GVD of the designed tapered waveguide. **(d)** Net TOD of the designed tapered waveguide.

Figure S2a shows the high resolution swept wavelength interferometer (SWI), which is applied for initial dispersion measurement. The microresonator transmission, from which FSR values are determined (Figure 2a inset), is measured using a tunable laser swept through its full wavelength tuning range at a tuning rate of 60 nm/s. For absolute wavelength calibration, 1% of the laser output was directed into a fiber coupled hydrogen cyanide gas cell (HCN-13-100, Wavelength References) and then into a photodetector ($PD_{REF}$). The microresonator and gas cell transmissions are recorded simultaneously during the laser sweep by a data acquisition system whose sample clock is derived from a high-speed photodetector ($PD_{MZI}$) monitoring the laser transmission through an unbalanced fiber Mach-Zehnder Interferometer (MZI). The MZI has a path length difference of approximately 40 m, making the measurement optical frequency sampling resolution 5 MHz. Figure S2b shows the transmission measurement of the SWI. The recorded resonances around the pump mode shows the critical coupling of our microresonator. It further proves that our tapered microresonator design not only engineers the dispersion, but also effectively suppresses transverse mode coupling.

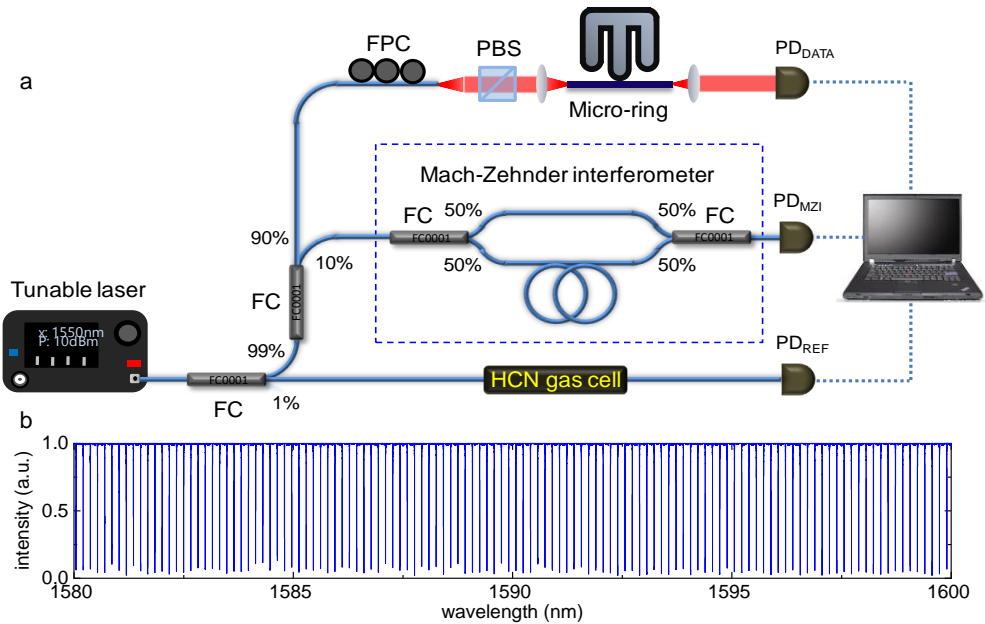

**Figure S2 | Device transmission dispersion characterization.** (a) Schematic setup for swept wavelength interferometer. (b) recorded resonances around the pump mode, showing no higher order mode coupling.

## II. Impact of Δ on the platicon generation with zero and positive TOD

In order to quantify the impact of the modulation frequency deviation $\Delta$ on the platicon microcomb, we introduce skewness, a measure to describe probability distribution of a real-valued random variable about its mean in statistics, to quantify the asymmetry of the comb line distribution. Here, we assign the comb line wavelength $\lambda_i$ as the variable, where the pump wavelength $\lambda_{pump}$ as the mean, and the value $P_i = \frac{I_i}{I_{total}}$ of the log-scale comb line intensity normalized to the total intensity as the probability for each comb line. Then the skewness is defined as: $skewness = \frac{\mu_3 - 3\mu\sigma^2 - \mu^3}{\sigma^3}$, where $\mu_3 = \sum_i \lambda_i^3 P_i$, $\mu = \sum_i \lambda_i P_i$, and $\sigma^2 = \sum_i \lambda_i^2 P_i - \mu^2$. In our cases, zero skewness indicates a perfectly symmetric comb spectrum, while negative skewness indicates more comb line power distributes on the longer wavelenght (lower frequency), and positive skew means more comb line power distributes on the shorter wavelength (higher frequency).

Figure S3(a-d) shows three simulated comb spectra of platicon at TOD $= 0$ fs$^3$/mm, with modulation frequency deviation $\Delta = -1000, 0,$ and $1000$ kHz, as well as the skewness, consistent with Figure 3. Note that the comb spectra here have frequency as the x axis, while the skewness is flipped for direct comparison with Figure 3g, and it is why the comb line distribution looks controdictary to the skewness plot. The skewness plot is central symmetric, and we could see that when $\Delta = 0$, skewness is zero, which indicates a perfect symmetric comb spectrum.

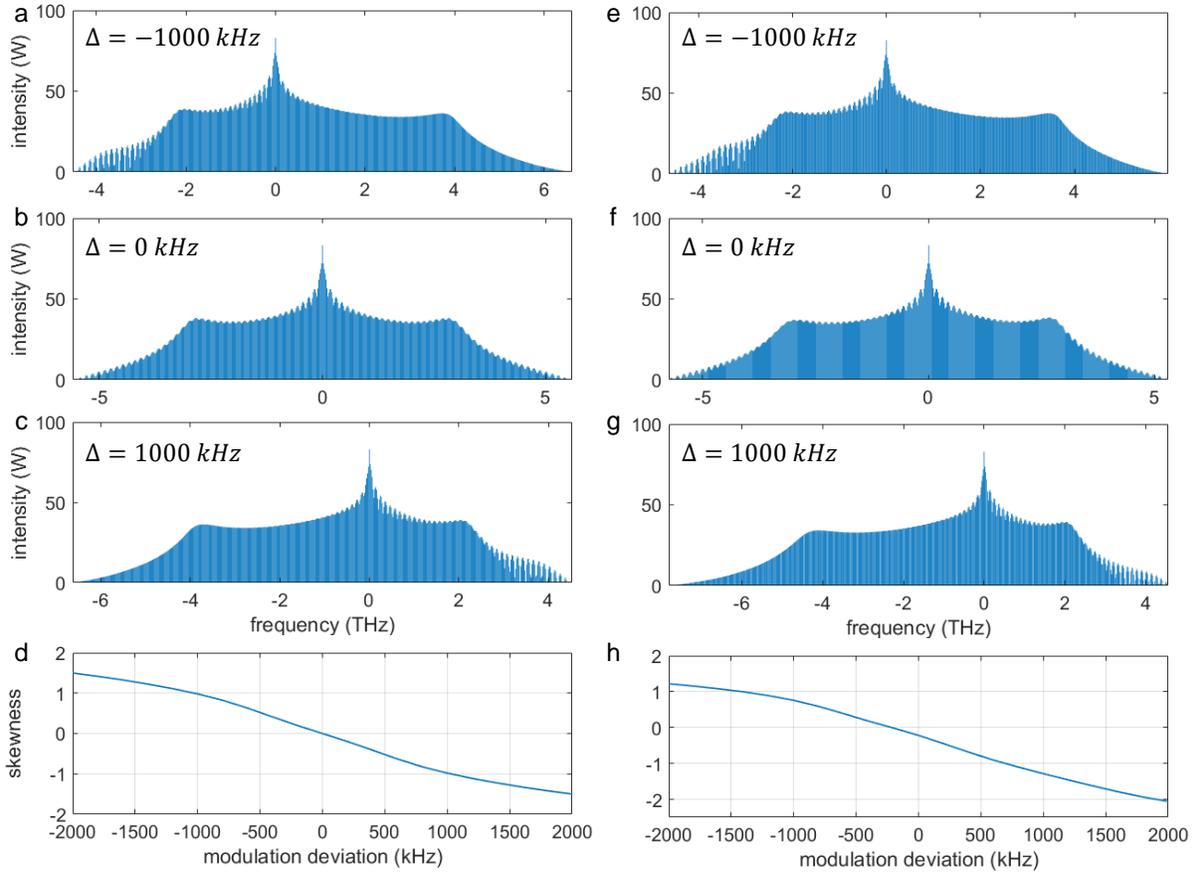

**Figure S3 | Modeled platicon comb spectra and calculated skewness for Δ = -1000 kHz, 0 kHz, and 1000 kHz when TOD = 0 and 1000 fs³/mm. (a-c)** modeled comb spectra for TOD = 0 fs³/mm. **(d)** Skewness for different Δ when TOD = 0 fs³/mm. **(e-g)** modeled comb spectra for TOD = 1000 fs³/mm. **(h)** Skewness for different Δ when TOD = 1000 fs³/mm.

When the TOD remain the small amount as the real case introduced in the main text, but with negative sign (TOD = 1000 fs³/mm), the phenomenon is inverted compared to the negative case (Figure 3). The platicon comb is asymmetric to longer wavelength (lower frequency) when $\Delta = 0$, and the comb spectra will be more asymmetric when Δ is positive, comparing to the negative case.

In the experiment, we observe that the broadband platicon comb could only be achieved within certain range of modulation frequency, shown in Figure S4. When the modulation frequency is within 19.54 GHz to 19.556 GHz, the comb spectra still maintain the signature of a platicon comb, however, the modulation frequency is outside these numbers, the comb spectra become much narrower, and most of the comb lines are buried by the residual ASE noise of EDFA.

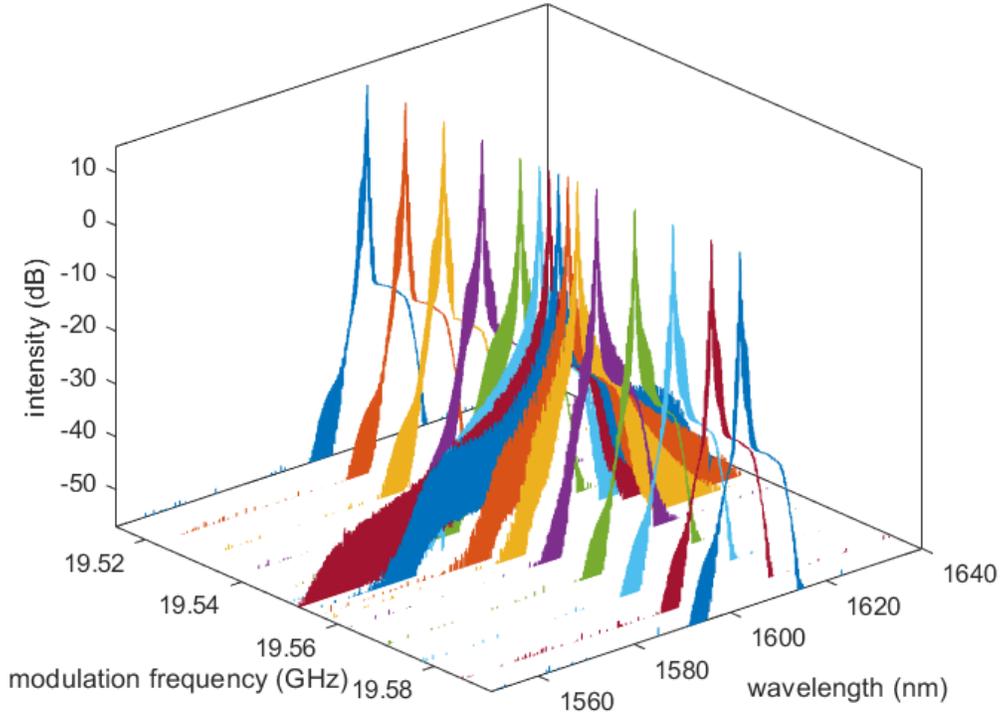

**Figure S4 | experimental comb spectra at different modulation frequency.** The broadband platicon comb is only experimentally achievable within certain modulation frequency. Otherwise, only narrowband triangle shape comb is observable.

## III. Noise performance characterization of the platicon

Since the comb spacing is intrinsically locked to the modulation frequency and there are no sub-comb families, the amplitude noise remains intrinsically low, and the electrical beat note remains clean with high signal-to-noise-ratio (SNR) throughout the whole evolution. Figure S5 shows the amplitude noise and heterodyne beatnote of the repetition rate at RBW of 10 Hz. The beat note power spectral density and amplitude noise is solely instrument limited, and the beat note has a clear Lorentzian lineshape.

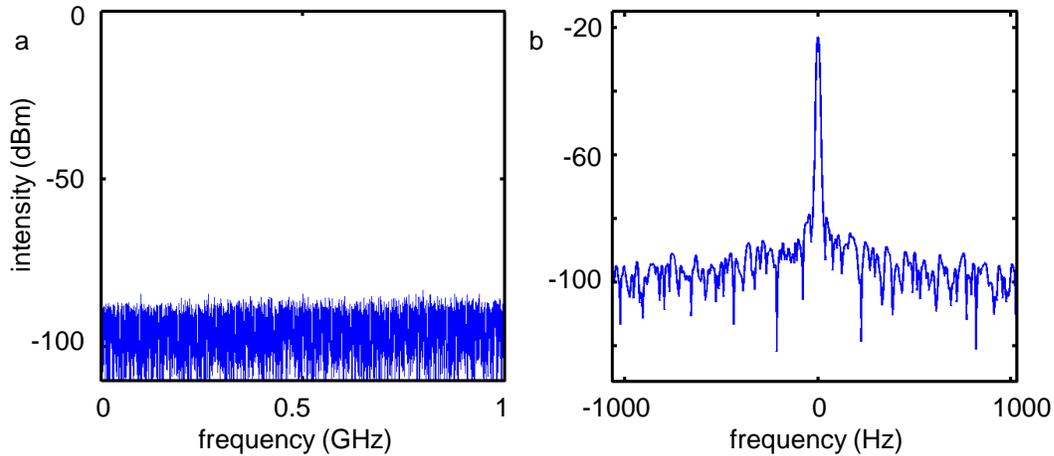

**Figure S5 | noise performance characterization.** (a) amplitude noise measurement from DC to 1 GHz. (b) Heterodyne beatnote of the repetition rate at RBW of 10 Hz.

## IV. Dual-comb cross-correlation

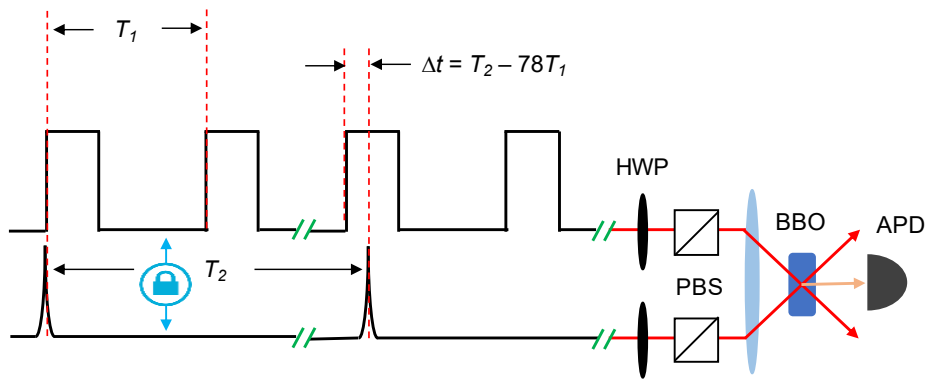

**Figure S6 | Schematic setup for dual-comb cross-correlation.** An ultrashort pulse with repetition rate of 250 MHz is utilized to probe the flat-top square pulse, and then the convolution is achieved through SHG.

    The pump modulation frequency (and thus the platicon frequency comb spacing) and the Menlo fiber reference laser comb spacing are both referenced to the same Rb-disciplined crystal oscillator. By tuning the Menlo fiber laser comb spacing close to 78 times of the modulation frequency, dual-comb cross-correlation can be achieved due to the temporal walk-off between the picosecond platicon and the femtosecond reference pulse. As shown in Figure S6, the two locked pulse trains will have a time lag $\Delta t = T_2 - 78T_1$, after every 78 cycles of the square pulse train. With both stable pulses trains sent into two half-wave plates (HWPs) and polarization beam splitters (PBS) and then aligned into a bulk barium borate (BBO) for second-harmonic generation, the cross-correlation sampled signal is collected by a 700 Hz femtowatt avalanche photodiode

(APD). A high-speed oscilloscope is then used to record the platicon temporal shape. Figure 4b of the main text shows the observed 17 ps flat-top pulse train. The rising edge has a 10-to-90% rise time of 5.6 ± 0.48 ps, and the falling edge has a 90-to-10% fall time of 9.54 ± 0.82 ps, sampled over 100 pulses. Notably, such rising and falling time could not precisely reflect the real scenario. The resolution of the cross-correlation method is determined by the pulse width of the Menlo fiber laser comb. However, in order to suppress the nonlinear effect of the Menlo comb itself in the cross-correlation, the fs fiber laser is pre-chirped, which would substantially reduce the resolution of the cross-correlation method. Hence, instead of precisely measure the square pulse, the measurements here only qualitatively characterize the flat-top feature and the pulse width of the platicon pulse. The numerical modelled 17 ps platicon pulse reveals that the rising and falling edges to be 0.31 ps and 0.7 ps, respectively.

**Supplementary Information Section V: Auxiliary-laser-assisted thermal stabilization**

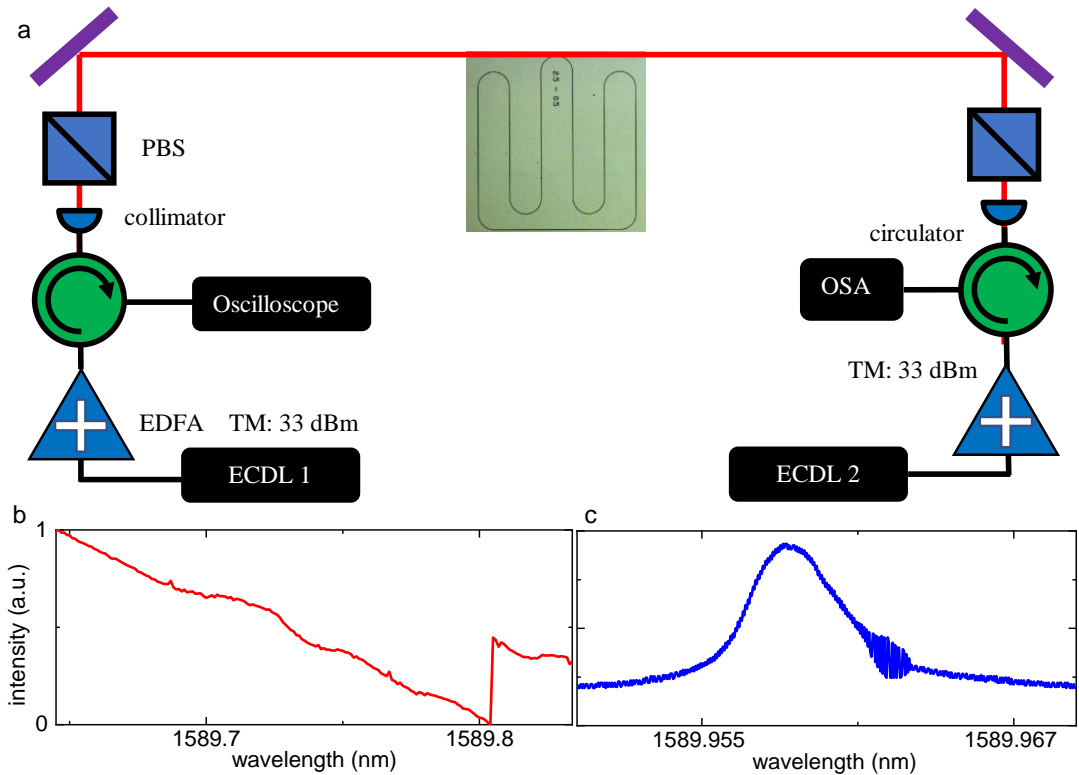

**Figure S7 | Measurement schematic for dual-comb cross-correlation.** (**a**) Schematic setup of platicon generation in normal dispersion regime using dual-driven method. (**b**) pump power transmission across a resonance without dual-driven method. (**c**) pump power transmission across a resonance with dual-driven method. Effective red detuning is accessed.

Due to the thermal nonlinearity, as shown in Figure S7b, the effective red detuning becomes so sharp that it is challenging to stop pump frequency at red detuning side. The red detuning studies for the platicon comb would complement the blue detuning studies. In order to access to more dynamics of platicon at red detuning side, we introduce dual-driven method to thermally stabilize the resonance. By sending an auxiliary pump backwardly into a resonance in C-band, the microresonator is heated. Then a single-FSR intensity modulated pump is tuned into a resonance in L-band in a forward direction. Then the resonance further red shifts, pushing the aux pump out of the resonance, effectively cooling the microresonator, and vice versa. Such balance effectively achieves thermal stabilization of the resonance, making red detuning accessible. Figure 1e shows the auxiliary pump transmission, showing accessible red detuning side. The oscillation on the left is due to the interference between the comb line generated by the IM pump and the comb line generated by the auxiliary pump via XPM.

In this work we experimentally examine the platicon generation with modulated pump at single FSR due to its practicability in real experimental environment. Platicons have also been studied numerically under different scenarios [1–5] which offer valuable insights to the dynamics of platicons.